\documentclass[letterpaper,conference]{ieeeconf}
\usepackage{graphicx}
\usepackage{cite}
\usepackage{epstopdf}
\usepackage{amsmath}
\usepackage{bm}
\usepackage{subfigure}
\usepackage{array,booktabs}
\usepackage{multirow}
\usepackage{tabularx}
\usepackage{threeparttable}
\usepackage[table]{xcolor}
\newcommand{\cellc}{\cellcolor{blue!10}}

\IEEEoverridecommandlockouts
\begin{document}

\title{\LARGE \bf Time-Series Prediction of Proximal Aggression Onset in Minimally-Verbal Youth with Autism Spectrum Disorder Using Physiological Biosignals}

\author{Ozan \"{O}zdenizci, Catalina Cumpanasoiu, Carla Mazefsky, Matthew Siegel,\\ Deniz Erdo\u{g}mu\c{s}, Stratis Ioannidis, Matthew S. Goodwin%
\thanks{O.~\"{O}zdenizci, C.~Cumpanasoiu, D.~Erdo\u{g}mu\c{s}, S.~Ioannidis and M.~S.~Goodwin are with Northeastern University, Boston, MA, USA. M.~Siegel is with Maine Medical Center Research Institute, Portland, ME, USA. C.~Mazefsky is with University of Pittsburgh, Pittsburgh, PA, USA. Corresponding author: oozdenizci@ece.neu.edu.}%
}

\maketitle


\begin{abstract}
It has been suggested that changes in physiological arousal precede potentially dangerous aggressive behavior in youth with autism spectrum disorder (ASD) who are minimally verbal (MV-ASD). The current work tests this hypothesis through time-series analyses on biosignals acquired prior to proximal aggression onset. We implement ridge-regularized logistic regression models on physiological biosensor data wirelessly recorded from 15 MV-ASD youth over 64 independent naturalistic observations in a hospital inpatient unit. Our results demonstrate proof-of-concept, feasibility, and incipient validity predicting aggression onset 1 minute before it occurs using global, person-dependent, and hybrid classifier models.
\end{abstract}



\IEEEpeerreviewmaketitle

\section{Introduction}

A substantial number of youth with autism spectrum disorder (ASD) show unpredictable and potentially dangerous aggressive behavior \cite{Kanne:2011,Farmer:2011,Gray:2012,Woodman:2016}. Aggression is particularly impairing in youth with ASD who are minimally verbal (MV-ASD). Due to their difficulty to adequately self-report increasing distress, aggression in MV-ASD is often unpredictable, which makes it dangerous, difficult to manage, and constitutes a barrier to accessing the community. In typically developing youth, greater ability to regulate physiological arousal is associated with fewer behavior problems \cite{Porges:1996}. Prior research suggests an association between physiological arousal and problem behavior to alleviate distress in ASD \cite{Groden:1994,Groden:2005,Cohen:2011,Romanczyk:1992,Groden:2006}. Thus, our current study evaluates whether proximal biomarkers of physiological arousal \cite{Goodwin:2006,Benevides:2015} can be used to predict the onset of aggression before it occurs.

A growing number of researchers have approached this goal using various observational designs \cite{Liu:2008,Kushki:2013,Kushki:2015,Welch:2012,Lydon:2013}; however, they all rely on artificial experimental settings and tasks that call into question the ecological validity of obtained results. In the present study, we utilize a novel data set in which physiological biosignals were collected wirelessly from behaviorally unstable MV-ASD youth during unstructured inpatient hospital observations. We employ ridge-regularized linear classifiers with time-series features of cardiovascular and electrodermal activity as physiological biomarkers of imminent aggression. Using global, person-dependent, and hybrid (i.e., partially global) models, we demonstrate high predictability of aggressive behavior onset in a specific upcoming time interval. A preliminary version of this work was presented in \cite{Ozdenizci:2018}.

\section{Materials and Methods}

\subsection{Participants and Data Acquisition}
\label{daq}

Physiological and physical activity data collected from $15$ ADOS-2 \cite{Lord:2012} confirmed, behaviorally unstable MV-ASD youth recorded over $64$ unstructured observational sessions in the Developmental Disorders Unit at Spring Harbor Hospital, Portland, ME, United States, were used in this Institutional Review Board approved study. Data were collected during naturalistic observational sessions at this specialized ASD psychiatric inpatient unit using a wrist-worn E4 biosensor (Empatica Inc., United States) as well as time-synchronized and operationally defined (hitting, kicking, biting, scratching, grabbing, pulling) coding of aggression to others by inpatient research staff with at least $90\%$ inter-rater reliability. All MV-ASD youth in our sample tolerated the biosensor after desensitization, usable data were obtained in all cases, and a total observation time of approximately $83$ hours was achieved with an average of $5.5\pm4.9$ hours across participants. Aggressive episodes were observed $35.9$ (SD $=34.1$) times with average durations of $31.9\pm33.2$ seconds.

To capture measures of physiological arousal, the following autonomic nervous system indices were recorded by the wrist-worn E4 biosensor: (1) heart rate and heart rate variability, which is a measure of the variation in beat-to-beat interval, both derived from \textit{blood volume pulse} ($BVP$) and \textit{inter-beat interval} ($IBI$) via photoplethysmography \cite{Allen:2007} at $64$ Hz; and (2) \textit{electrodermal activity} ($EDA$) sampled at $4$ Hz, which reflects autonomic innervation of sweat glands and provides a sensitive measure of alterations in physiological arousal. To quantify changes in physical activity, the E4 records \textit{movement acceleration} ($ACC$) using an embedded $3$-axis accelerometer at $32$ Hz sampling rate.

\subsection{Time-Series Feature Extraction}
\label{featextraction}

Statistical analyses of physiological and physical activity signals were performed through extracted time-series features offline. For each one of the $6$ signal sources (i.e., $BVP$, $IBI$, $EDA$, $ACC_x$, $ACC_y$, $ACC_z$) the following features were calculated in bins of $15$ seconds: first, last, maximum, minimum, mean and median value, amount of unique values, and the sum, standard deviation, and variance of values within a bin. To exploit temporal information of aggression episodes, we extracted two more features using time-synchronized binary aggression labels offline; \textit{time since past aggression} ($TPA$), which indicates the amount of time elapsed since the last observation of an aggression episode; and a binary \textit{aggression observation flag} ($AOF$) feature, indicating whether an aggression episode has so far occurred within that recording session. The standard deviation of each calculated feature across time-series bins was included in all prediction models.

\subsection{Ridge-Regularized Logistic Regression Classifier}
\label{classifierframework}

Binary decision making on aggressive behavior onset was performed through a ridge-regularized logistic regression model. Among all extracted features (see Section~\ref{featextraction}), the following five different feature subsets were used as predictor variables in our analyses: (1) only temporal information ($TPA$, $AOF$); (2) only physical activity ($ACC$); (3) only physiological activity ($BVP$, $IBI$, $EDA$); (4) physical and physiological activity features combined ($BVP$, $IBI$, $EDA$, $ACC$); and (5) all features combined ($BVP$, $IBI$, $EDA$, $ACC$, $TPA$, $AOF$).

At every time point $\textit{t}$, using features extracted in a previous time range $[\textit{t}-\tau_p, \textit{t})$, a classifier was used to predict a binary dependent variable, estimating whether aggression will be observed or not in an upcoming time range $(\textit{t}, \textit{t}+\tau_f]$. We adopted a $5$-fold cross-validation protocol to generate training and testing data splits, repeated five times to produce confidence intervals. Predictions with particular values of $\tau_p$ and $\tau_f$ were performed via the following schemes:

\begin{enumerate}
    \item \textit{Global Prediction Models:} Time-series data \textit{across} all participants and observational sessions were concatenated. Using the training data split obtained, a single classifier was constructed as the global model.
    \item \textit{Person-Dependent Prediction Models:} Data were pooled across all observational sessions \textit{within} each one of the $15$ participants individually. Obtained training data splits were used to construct $15$ distinct person-dependent prediction models.
    \item \textit{$k$-Hybrid Prediction Models:} In this framework, person-dependent prediction models were trained in a partially global scheme. Specifically, $k$ most significant physiological or physical biomarker features were identified via their regression weights in the global models (i.e., $k$ features with the highest weight magnitudes). Using each participant's pooled data, the corresponding $k$ feature weights were trained globally, whereas the other feature weights were trained solely on person-dependent data. In this scheme, $15$ $k$-hybrid prediction models are constructed for each participant.
\end{enumerate}

Considering a predictive variable feature space dimensionality of $d$ which varies with $\tau_p$, iterative training of logistic regression weights at every cross-validation fold requires estimation of $d$ hyperparameters for global models, and $d \times S$ hyperparameters for person-dependent models where $S$ denotes the number of participants. In $k$-hybrid models, the number of hyperparameters to be estimated are $((d-k) \times S) + k$. This becomes more computationally parsimonious with respect to person-dependent models when the number of participants $S$ is high.

\section{Results}

Classification analyses for particular values of $\tau_p$ and $\tau_f$ were performed through global, person-dependent, and $k$-hybrid models. To evaluate performance, Receiver Operator Characteristic (ROC) curves and corresponding Area Under the Curve (AUC) values were calculated through probabilities that logistic regression classifiers generate two classes: aggression and non-aggression. Data were processed for decision making every $15$ seconds, resulting in $18,869$ samples in global models.

\subsection{Global Prediction Models}
\label{globalresults}

Global prediction models of aggression in the upcoming one minute using all extracted features from the past $\tau_p = 60$ seconds resulted in the blue solid ROC curve presented in Figure~\ref{plots}(a) with a corresponding AUC value of $0.69$. Figure~\ref{plots}(b) depicts increases in AUC when E4 biosensor data is included, compared to using any other subset of features to predict aggression episodes.

Regarding the relationship between past time range ($\tau_p$) and future time range used to make aggression onset predictions ($\tau_f$), Figure~\ref{plots}(c) depicts stationary performance in global models using all features from various past $\tau_p$ durations. However, when using physical and physiological biosensor data from the past (c.f.~bold dashed red curve), we observe relative AUC increases compared to temporal data of past aggression only (c.f.~light dotted dark blue curve).

\subsection{Person-Dependent Prediction Models}

\begin{figure*}[th!]
	\centering
	\subfigure[]{\includegraphics[height=0.346\textwidth]{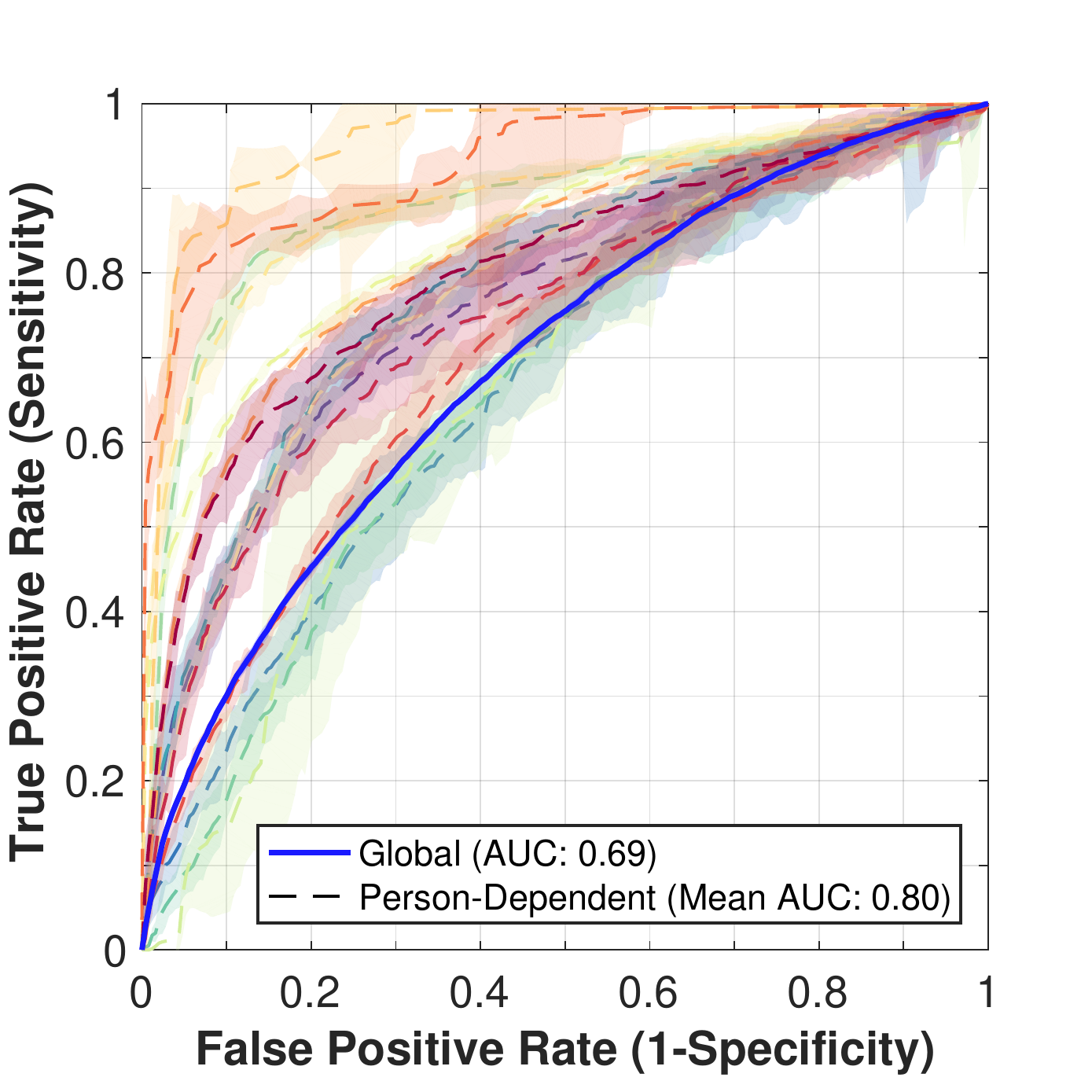} \hspace{-0.4cm}}
	\subfigure[]{\includegraphics[height=0.34\textwidth]{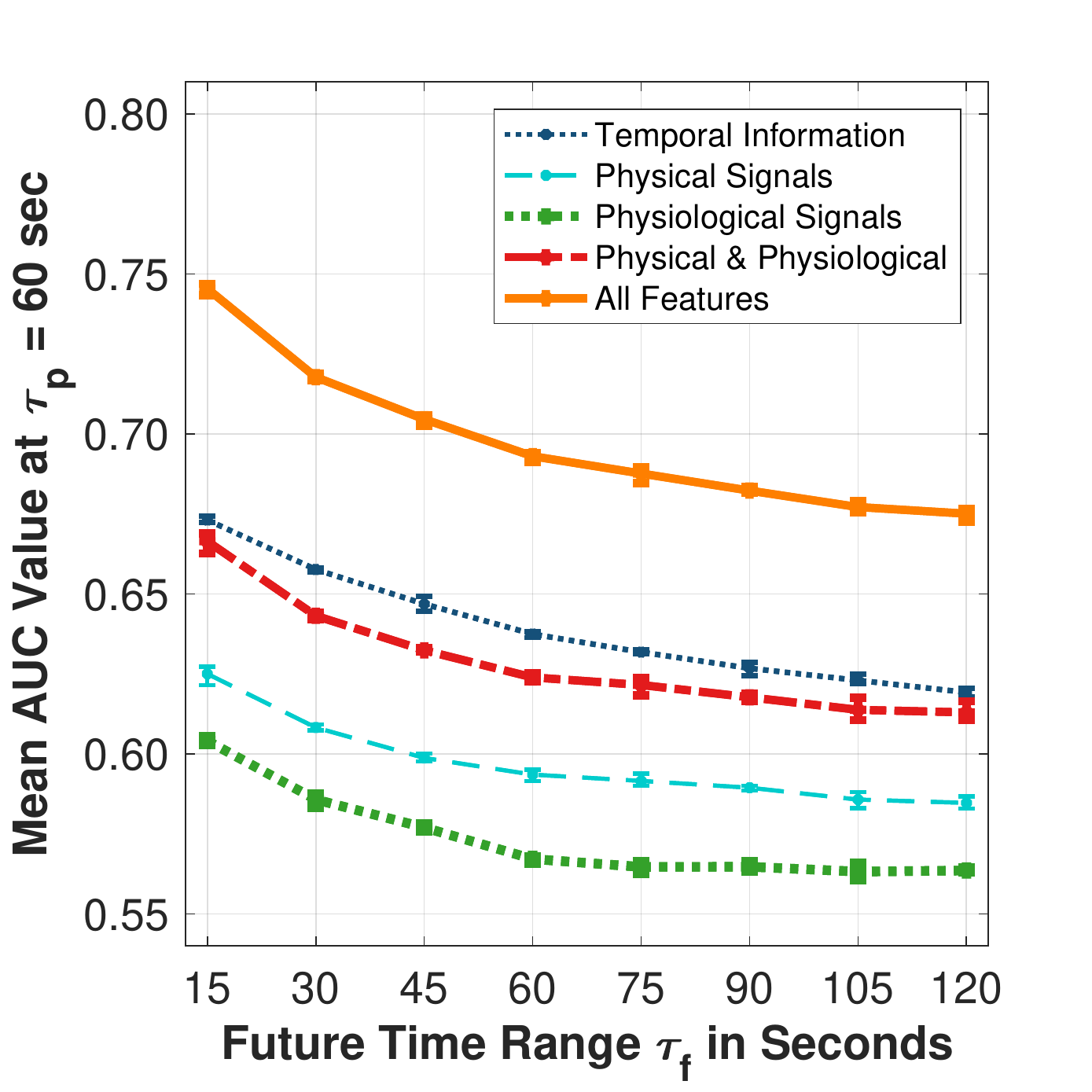} \hspace{-0.4cm}}
 	\subfigure[]{\includegraphics[height=0.34\textwidth]{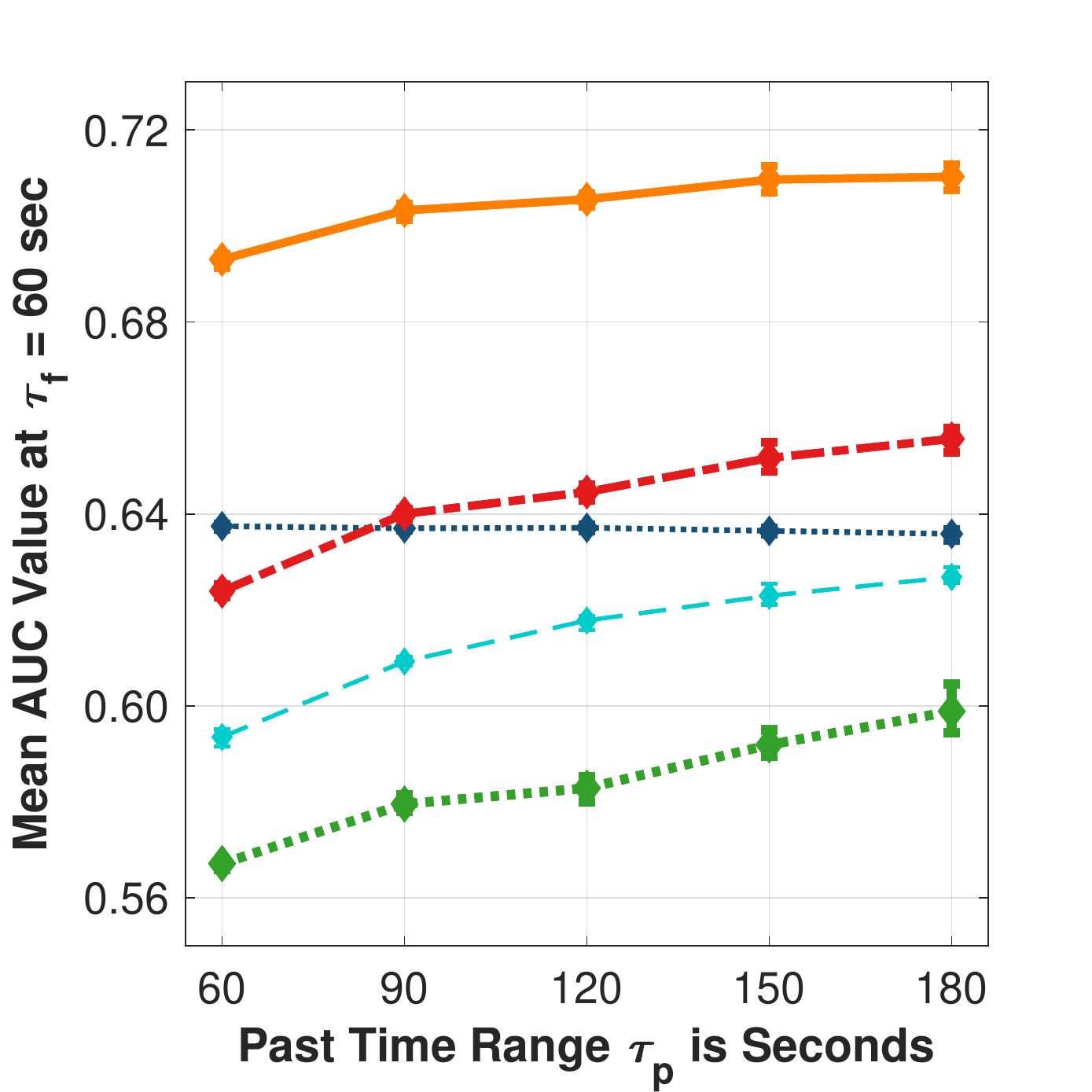}\hspace{-0.3cm}}
	\caption{(a) ROC curves with 90\% confidence intervals to predict onset of aggression in the next minute, using all features from the past minute. The blue solid line represents the global prediction model, and each curve with dashed lines represents one of the person-dependent models. (b) Mean AUC values of the global models varying as a function of $\tau_f$ using features from the past minute; and (c) as a function of $\tau_p$ at $\tau_f = 60$ seconds. For (b) and (c), each color represents one of the five feature subsets as indicated. Error bars represent minimum and maximum values across cross-validation repetitions.}
	\label{plots}
\end{figure*}

We repeat the above analysis for $\tau_p = \tau_f = 60$ seconds in person-dependent models; Table~\ref{persondeptable} contains the corresponding mean AUCs. Across participants, we observe an average $0.16$ increase in AUC values up to $0.80$ using E4 biosensor data compared to using temporal aggression information only. Figure~\ref{plots}(a) shows ROC curves corresponding to person-dependent model prediction performance using all features compared to the global prediction model with the same parameters. We observe a higher mean AUC and a much favored behavior of steep increases of sensitivity for low false positive rates in most person-dependent models compared to global prediction models.

\begin{table*}
  \vspace{1.1cm}
  \centering
  \caption{AUC values averaged across cross-validation repetitions in person-dependent prediction models with $\tau_p = \tau_f = 60$ seconds  (i.e., predicting aggression onset for the next minute, using accumulated data from the past one minute).}
  \label{persondeptable}
  \renewcommand{\arraystretch}{1.95}
  \scalebox{0.85}{
	\begin{tabular} {>{\centering\bfseries}m{0.86in} >{\centering}m{0.23in} >{\centering}m{0.23in} >{\centering}m{0.23in} >{\centering}m{0.23in} >{\centering}m{0.23in} >{\centering}m{0.23in} >{\centering}m{0.23in} >{\centering}m{0.23in} >{\centering}m{0.23in} >{\centering}m{0.23in} >{\centering}m{0.23in} >{\centering}m{0.23in} >{\centering}m{0.23in} >{\centering}m{0.23in} >{\centering}m{0.23in} >{\centering\arraybackslash}m{0.33in} >{\centering\arraybackslash}m{0.29in}}
    \toprule
    Features & \textbf{P1} & \textbf{P2} & \textbf{P3} & \textbf{P4} & \textbf{P5} & \textbf{P6} & \textbf{P7} & \textbf{P8} & \textbf{P9} & \textbf{P10} & \textbf{P11} & \textbf{P12} & \textbf{P13} & \textbf{P14} & \textbf{P15} & \textbf{Mean} & \textbf{SD} \\
    \midrule
	Temporal & 0.67 & 0.58 & 0.63 & 0.66 & 0.77 & 0.54 & 0.67 & 0.68 & 0.73 & 0.59 & 0.65 & 0.51 & 0.61 & 0.65 & 0.63 & \cellc\textbf{0.64} & \cellc\textbf{0.07} \\
	Physical & 0.63 & 0.69 & 0.69 & 0.62 & 0.76 & 0.53 & 0.76 & 0.71 & 0.80 & 0.76 & 0.77 & 0.92 & 0.66 & 0.55 & 0.75 & \cellc\textbf{0.71} & \cellc\textbf{0.10} \\
	Physiological & 0.76 & 0.57 & 0.76 & 0.56 & 0.82 & 0.61 & 0.80 & 0.61 & 0.83 & 0.90 & 0.77 & 0.86 & 0.58 & 0.65 & 0.76 & \cellc\textbf{0.72} & \cellc\textbf{0.11} \\
	Physical and Physiological & 0.75 & 0.67 & 0.76 & 0.63 & 0.84 & 0.63 & 0.83 & 0.73 & 0.85 & 0.95 & 0.81 & 0.93 & 0.69 & 0.74 & 0.80 & \cellc\textbf{0.77} & \cellc\textbf{0.10} \\
	All Features Combined & 0.76 & 0.67 & 0.79 & 0.67 & 0.88 & 0.67 & 0.85 & 0.79 & 0.89 & 0.96 & 0.82 & 0.93 & 0.70 & 0.75 & 0.80 & \cellc\textbf{0.80} & \cellc\textbf{0.09} \\
    \bottomrule
  \end{tabular}}
\end{table*}

\begin{table*}
  \vspace{1.1cm}
  \centering
  \caption{AUC values averaged across cross-validation repetitions in prediction models with $\tau_p = \tau_f  = 60$ seconds. By definition of the $k$-hybrid models, $k=0$ corresponds to person-dependent models, whereas $k=310$ corresponds to the global model.}
  \label{hybridtable}
  \renewcommand{\arraystretch}{1.9}
  \scalebox{0.85}{
	\begin{tabular} {>{\centering\bfseries}m{0.64in} >{\centering}m{0.22in} >{\centering}m{0.23in} >{\centering}m{0.23in} >{\centering}m{0.23in} >{\centering}m{0.23in} >{\centering}m{0.23in} >{\centering}m{0.23in} >{\centering}m{0.23in} >{\centering}m{0.23in} >{\centering}m{0.23in} >{\centering}m{0.23in} >{\centering}m{0.23in} >{\centering}m{0.23in} >{\centering}m{0.23in} >{\centering}m{0.23in} >{\centering}m{0.23in} >{\centering\arraybackslash}m{0.33in} >{\centering\arraybackslash}m{0.29in}}
    \toprule
    Model & \textbf{$k$} & \textbf{P1} & \textbf{P2} & \textbf{P3} & \textbf{P4} & \textbf{P5} & \textbf{P6} & \textbf{P7} & \textbf{P8} & \textbf{P9} & \textbf{P10} & \textbf{P11} & \textbf{P12} & \textbf{P13} & \textbf{P14} & \textbf{P15} & \textbf{Mean} & \textbf{SD} \\
    \midrule
    Person-Dependent & 0 & 0.76 & 0.67 & 0.79 & 0.67 & 0.88 & 0.67 & 0.85 & 0.79 & 0.89 & 0.96 & 0.82 & 0.93 & 0.70 & 0.75 & 0.80 & \cellc\textbf{0.80} & \cellc\textbf{0.09} \\
	$k$-Hybrid & 1 & 0.73 & 0.58 & 0.74 & 0.60 & 0.86 & 0.57 & 0.83 & 0.76 & 0.86 & 0.87 & 0.81 & 0.87 & 0.67 & 0.69 & 0.75 & \cellc\textbf{0.75} & \cellc\textbf{0.11} \\
	$k$-Hybrid & 10 & 0.74 & 0.57 & 0.75 & 0.62 & 0.86 & 0.54 & 0.83 & 0.76 & 0.87 & 0.86 & 0.81 & 0.87 & 0.67 & 0.69 & 0.74 & \cellc\textbf{0.75} & \cellc\textbf{0.11} \\
	$k$-Hybrid & 100 & 0.75 & 0.57 & 0.76 & 0.67 & 0.87 & 0.53 & 0.83 & 0.77 & 0.86 & 0.85 & 0.81 & 0.83 & 0.69 & 0.71 & 0.73 & \cellc\textbf{0.75} & \cellc\textbf{0.10} \\
	$k$-Hybrid & 200 & 0.76 & 0.59 & 0.73 & 0.70 & 0.87 & 0.52 & 0.83 & 0.77 & 0.87 & 0.84 & 0.80 & 0.83 & 0.68 & 0.71 & 0.72 & \cellc\textbf{0.75} & \cellc\textbf{0.10} \\
	$k$-Hybrid & 300 & 0.74 & 0.60 & 0.70 & 0.69 & 0.85 & 0.54 & 0.81 & 0.75 & 0.83 & 0.70 & 0.73 & 0.77 & 0.66 & 0.67 & 0.62 & \cellc\textbf{0.71} & \cellc\textbf{0.09} \\
	Global & 310 & $-$ & $-$ & $-$ & $-$ & $-$ & $-$ & $-$ & $-$ & $-$ & $-$ & $-$ & $-$ & $-$ & $-$ & $-$ & \cellc\textbf{0.69} & \cellc$-$ \\
    \bottomrule
  \end{tabular}}
\end{table*}

\subsection{$k$-Hybrid Prediction Models} 

Table~\ref{hybridtable} demonstrates results obtained by $k$-hybrid models with $\tau_p = \tau_f = 60$ seconds for varying values of $k$. For parameter $\tau_p = 60$, dimensionality of feature vectors became $d = 310$ when all extracted features in bins of 15 seconds were concatenated to be used as predictor variables. In this case, by definition (see Section \ref{classifierframework}), a $k$-hybrid model with $k = 310$ is equivalent to a global prediction model. Mean AUC values of the $k$-hybrid models yielded stationary performance between person-dependent and global models with $0.75$ for a $k$ value as high as $200$. While partially globalizing the model slightly reduced AUC values compared to person-dependent models, model training becomes computationally more efficient. Thus, the $k$-hybrid framework serves as a partially global model alternative that approximates person-dependent accuracies with shorter model training run time. Furthermore, although person-dependent models provide better accuracies, they cannot be used to predict aggression onset for a new patient for whom no prior data is available. From that perspective, hybrid models are desirable in that predictive power can be maintained by having some parameters trained globally, and then combined and updated as person-dependent training data becomes available.

\section{Discussion}

In the present study we demonstrate that naturalistically observed aggressive behavior in MV-ASD youth in a hospital inpatient setting can be predicted with high accuracy using preceding physiological and physical activity biosensor data and temporal information on recently observed aggressive episodes. We implemented linear classifier models with regularization over relevant time-series features during training. Our results demonstrate proof-of-concept, feasibility, and incipient validity predicting proximal aggression in this population and setting using biosignal data in global, person-dependent, and hybrid models.

As communicated to us by inpatient clinical staff, the potential benefit of avoiding or reducing a dangerous aggressive event is likely to outweigh the potential harm associated with a false positive in clinical practice. Predicting aggression onset during naturalistic observation in the upcoming $1$ minute with at least $80\%$ sensitivity is of high clinical value and could create new opportunities for preventing or mitigating aggression emergence, occurrence, and impact in MV-ASD \cite{Ozdenizci:2018}.

Regarding future work, we will next model aggression onsets as a non-homogeneous Poisson process to assess whether they can yield more robust global prediction models. Using this framework, we will also evaluate whether regressing hazard rates from past observations can be performed through maximum likelihood estimation, comprising an iterative solution of several weighted linear regressions \cite{Massey:1996}, and lead to improved predictive performance across upcoming time ranges.

\section*{Acknowledgment}

This work was supported by grants from the Simons Foundation (SFARI 296318), the Nancy Lurie Marks Family Foundation, the National Institute on Deafness and Other Communication Disorders (P50 DC013027), and NSF (SCH-1622536, IIS-1118061).




\end{document}